\documentclass[10pt,twocolumn]{article}

\usepackage{natbib}
\usepackage{graphicx}

\usepackage{times}

\renewcommand\refname{References and Notes}
   \newcommand{\kms}{{km\,s$^{-1}~$}}
     \newcommand{\halpha}{H$\alpha$}
      \newcommand{\arcsec}{''}

\newcounter{lastnote}
\newenvironment{scilastnote}{%
\setcounter{lastnote}{\value{enumiv}}%
\addtocounter{lastnote}{+1}%
\begin{list}%
{\arabic{lastnote}.}
{\setlength{\leftmargin}{.22in}}
{\setlength{\labelsep}{.5em}}}
{\end{list}}

\title{Measuring the cosmic ray acceleration efficiency of a supernova remnant} 

\author
{E.A.Helder $^{1\diamond}$, J. Vink $^{1}$, C.G. Bassa$^{2,3}$, A. Bamba$^{4}$, J.A.M. Bleeker$^{1,2}$, S. Funk$^{5}$, \\P. Ghavamian$^{6}$, K. J. van der Heyden$^{7}$, F. Verbunt$^{1}$ and R. Yamazaki$^{8}$ \\
\\
{\small $^{1}$Astronomical Institute Utrecht, Utrecht University, P.O. Box 80000 NL-3508 TA, Utrecht, The Netherlands}\\
{\small $^{2}$SRON Netherlands Institute for Space Research, Sorbonnelaan 2, NL-3584 CA Utrecht, The Netherlands}\\
{\small $^{3}$Department of Astrophysics, Radboud University Nijmegen, P.O. Box 9010, Nijmegen, The Netherlands}\\
{\small $^{4}$ISAS/JAXA Department of High Energy Astrophysics 3-1-1, Yoshinodai, Sagamihara, Kanagawa 229-8510, Japan}\\
{\small $^{5}$Kavli Institute for Particle Astrophysics and Cosmology, Stanford, CA-94025, USA}\\
{\small $^{6}$Space Telescope Science Institute, 3700 San Martin Drive, Baltimore, MD, 21218, USA}\\
{\small $^{7}$Astronomy Department, University of Cape Town, Private Bag X3, Rondebosch 7701, South Africa}\\
{\small $^{8}$Department of Physical Science, Hiroshima University, Higashi-Hiroshima, Hiroshima 739-8526, Japan}\\
\\
{\small $^\diamond$To whom correspondence should be addressed; E-mail:  e.a.helder@uu.nl.}
}

\date{}

\begin{document}


\baselineskip12pt


\maketitle

\def\figurename{Fig.}
\long\def\@makecaption#1#2{%
  \vskip\abovecaptionskip
  \sbox\@tempboxa{#1: #2}%
  \ifdim \wd\@tempboxa >\hsize
{\small\sf    {\bf #1} #2}\par
  \else
    \global \@minipagefalse
    \hb@xt@\hsize{\hfil\box\@tempboxa\hfil}%
  \fi
  \vskip\belowcaptionskip}


{\bf 
Cosmic rays are the most energetic particles arriving at earth. Although most of them are thought to be accelerated by supernova remnants, the details of the acceleration process and its efficiency are not well 
determined. Here we show that the pressure induced by cosmic rays exceeds the thermal pressure behind the northeast shock of the supernova remnant RCW 86, where the X-ray emission is dominated by synchrotron radiation from ultra-relativistic electrons.  We determined the cosmic-ray content from the thermal Doppler broadening measured with optical spectroscopy, combined with a proper-motion study in X-rays. The measured post-shock proton temperature in combination with the shock velocity does not agree with standard shock heating, implying that $>$50\% of the post-shock pressure is produced by cosmic rays. 
}



The main candidates for accelerating cosmic rays up to at least $10^{15}$ eV are shell-type supernova remnants (SNRs), which are the hot, expanding plasma shells, caused by exploded stars (supernovas). In order to maintain the cosmic-ray energy density in the Galaxy, about 3 supernovae per century should transform 10 percent of their kinetic energy in cosmic-ray energy. Indeed, $\sim10^{14}$ eV electrons have been detected at forward shocks \cite{Koyama,Bamba} and possibly at reverse shocks \cite{Rho2002,Helder2008} of several shell-type remnants by their X-ray synchrotron emission, and particles with TeV energies have been detected in several SNRs by Cherenkov telescopes \cite{Aharonian2004,Albert}. 

\begin{figure*}[!t]
\begin{center}
\begin{tabular}{c c }
\includegraphics[angle = 0, width=0.43\textwidth]{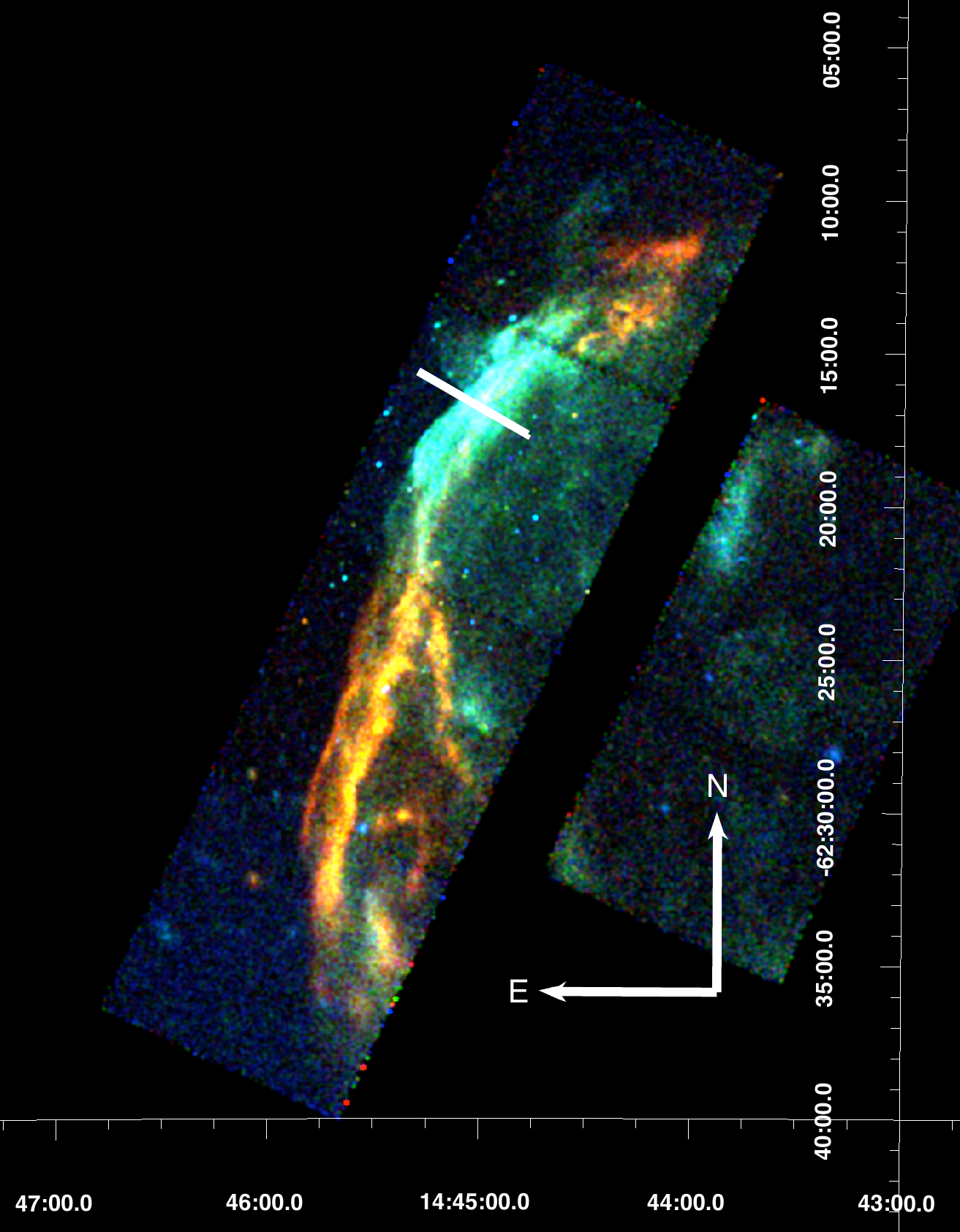} &
\includegraphics[angle = 0, width=0.42\textwidth]{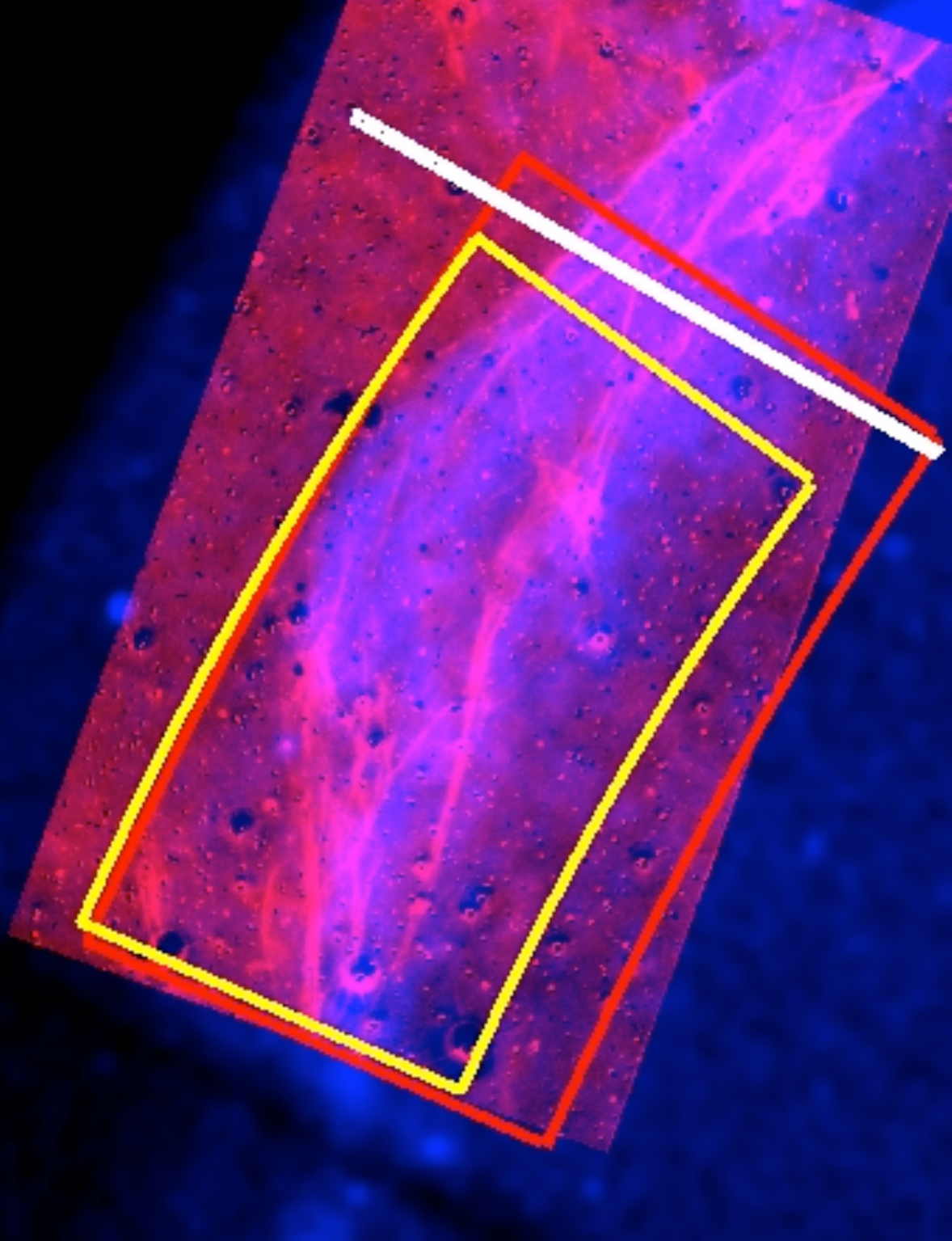}  \\
\end{tabular}
\caption {
({\bf Left}) The eastern rim of RCW~86, as observed in 2007 with Chandra
. Red indicates the 0.5-1.0 keV band, green the 1.0-1.95 keV band and blue shows the 1.95-6.0 keV band.  The northern part has relatively more flux in the higher energy bands, which is characteristic for synchrotron emission. ({\bf Right}) Blue is the broadband keV Chandra image, red is the image as observed with the VLT through a narrow H$\alpha$ filter. The regions (yellow and red) indicate where we measured the proper motion. 
In both panels the location where we took the optical spectrum is indicated with a white line.
}
              \label{Image}%
\end{center}
\end{figure*}

If SNRs transform a substantial amount of their kinetic energy into cosmic rays, this should affect the kinematics of the remnant. One imprint of energy losses by cosmic rays is a higher compression factor of the post-shock plasma\cite{Berezhko1999}, for which indications have been found in both the Tycho SNR and SN~1006 \cite{Warren2005,Cassam2008}. Another signature of the energy absorbed by cosmic rays is a lower post-shock temperature\cite{Decourchelle2000,Vinkreview2008, Drury2009,Patnaude2009b}. For shocks with conservation of mass, momentum and energy, in absence of cosmic rays, the post-shock temperature ($T_i$) for species with mass $m_i$ relates to the shock velocity ($v_s$) as 

\begin{equation} kT_i = \frac{3}{16}m_iv_s^2\label{heating}\end{equation}

in case of no thermal equilibrium (i.e. the several atomic species do not have the same temperature), in which case protons carry most of the thermal energy. In case of fast thermal equilibration, this relation reads $kT = \frac{3}{16}\mu m_{\rm p}v_s^2$ ($\mu \simeq0.6$ for cosmic abundances). Indications for a lower post-shock electron temperature have been found in the Magellanic Cloud remnant 1E 0102-72 \cite{Hughes0102}, which may constitute only a minor part of the thermal pressure. 
Here we derive the post-shock proton temperature and the shock velocity of the northeast rim of the shell-type SNR RCW~86 based on optical and X-ray observations. 

RCW~86 \cite{names} was detected in TeV energies by the H.E.S.S. telescope \cite{AharonianRCW} and is probably the remnant of the supernova witnessed by Chinese astronomers in 185 A.D.\cite{Stephenson,Vink2006}. It has been suggested that it evolves in a stellar-wind blown cavity, where the southwest corner has already hit the cavity wall\cite{Vink1997}. The northeast side still expands in a less dense medium and its X-ray spectrum is dominated by synchrotron radiation, which is an indication for efficient cosmic-ray acceleration. 

The optical spectrum of the northeast rim of RCW~86 is dominated by hydrogen lines, with no [NII] line emission\cite{Chevalier1980}. The lack of [NII] indicates that the hydrogen line emission is not a result of strong cooling, but results from excitation processes immediately behind the shock front. The hydrogen lines from these shocks consist of two superimposed Gaussian line profiles: one, caused by direct excitation, has the thermal width of the interstellar medium (ISM), the other is emitted after charge exchange between hot post-shock protons and cold incoming neutral hydrogen and hence has the thermal width of the post-shock protons. \halpha\ emission and efficient cosmic-ray acceleration are likely to anti-correlate because incoming neutral species are likely to damp plasma waves, which are essential for shock acceleration \cite{Drury1996} and because cosmic rays escaping ahead of the shock ionize the surrounding ISM and decrease the amount of \halpha\ emission. In RCW~86 the \halpha\ emission occurs all along the rim, including, although with weak emission, the parts coinciding with X-ray synchrotron emission, where efficient cosmic-ray acceleration is likely to occur (Fig. 1). The only other remnant in which \halpha\ emission is seen all along the shell, including regions with X-ray synchrotron emission, is the SN~1006 SNR\cite{Winkler2003}.

The right panel in Figure 1 shows both the H$\alpha$ and the X-ray emission of the northeast rim of RCW~86. The H$\alpha$ emission marks the onset of the X-ray synchrotron radiation, which indicates that they are from the same physical system. 

In order to measure the proton temperature, we used long-slit spectra obtained with the visual and near ultraviolet FOcal Reducer and low dispersion Spectrograph (FORS2) instrument on the Very Large Telescope (VLT) \cite{FORS}. We first imaged the northeast side of RCW~86, where the X-ray spectrum is dominated by synchrotron emission. Using this image as a guide, we pointed the slit at a location where the \halpha\ emission is bright (Fig. 1, Table S1). 

The spectrum's (Fig.2) measured full width at a half of the maximum (FWHM) is $1100 \pm 63$ km/s (see SOM for further details) corresponding to a $\sigma_v = 467\pm 27$ km/s and implying a post-shock temperature of 2.3 $\pm$ 0.3 keV.

\begin{figure}[!htb]
\begin{center}
\includegraphics[angle = 0, width=0.45\textwidth]{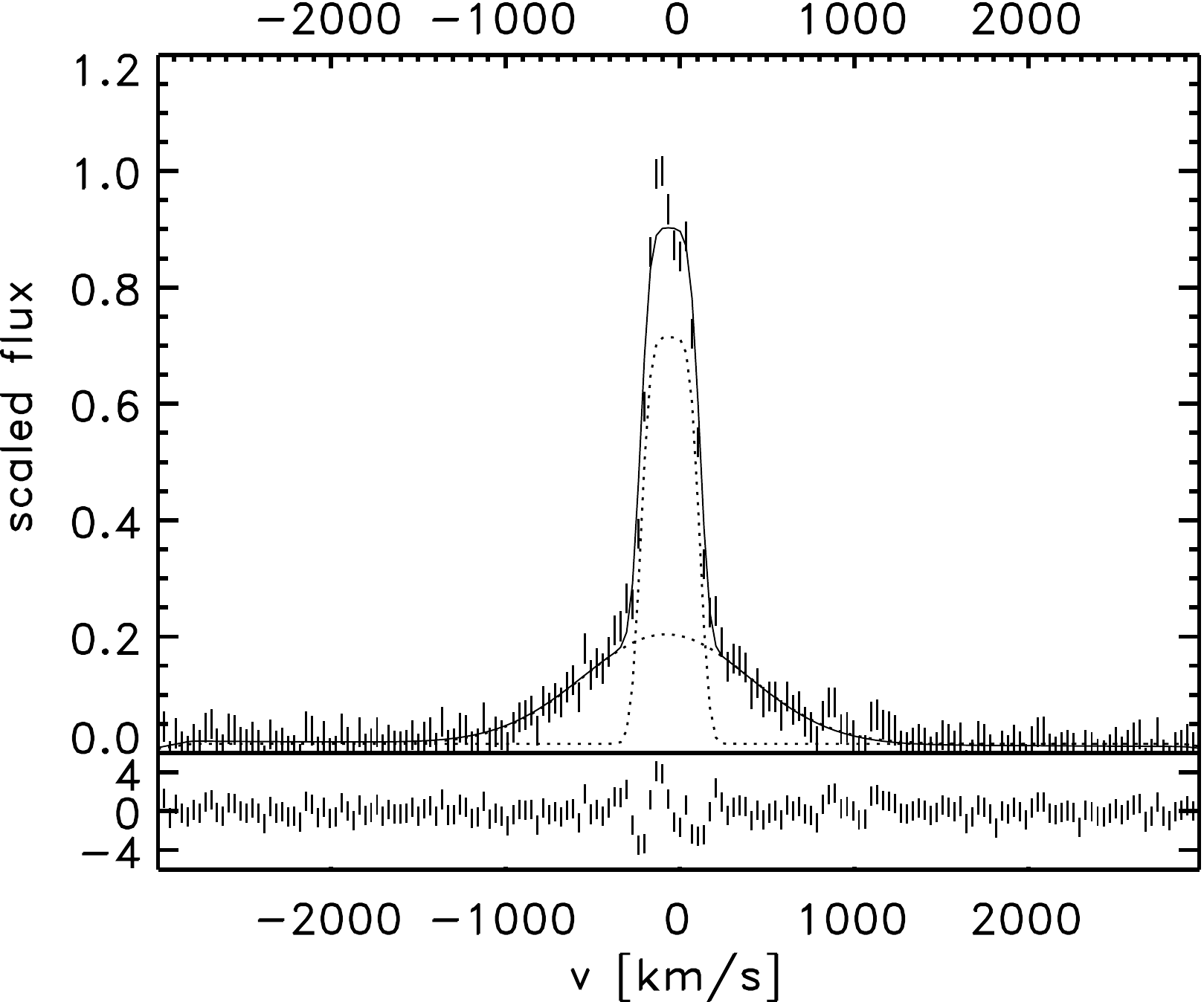} \\
\caption{
The H$\alpha$ spectrum, with broad and narrow components (dotted). The best fitting spectrum is overplotted. 
The lower panel shows the residuals divided by the errors.
}
              \label{Spectra}%
\end{center}
\end{figure}

To measure the shock velocity of the northeast rim of RCW 86, we observed it with the Chandra X-ray observatory in June 2007, and matched it with an observation taken in June 2004 \cite{Vink2006}. To make both observations as similar as possible, we used the same observation parameters as in 2004 (Table S2).

\begin{figure*}[!htb]
\begin{center}
\begin{tabular}{c c }
\includegraphics[angle = 0, width=0.42\textwidth]{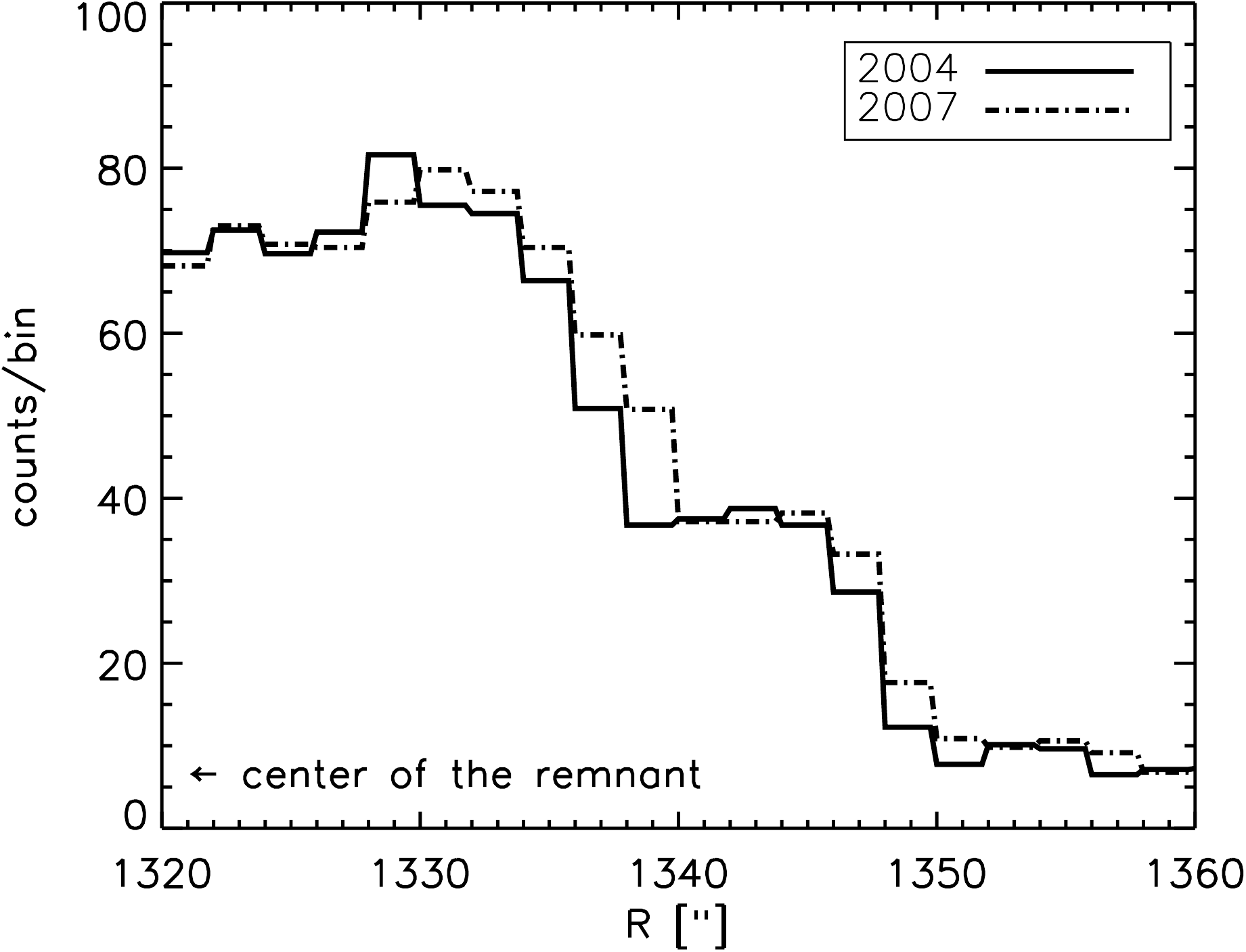}%
\includegraphics[angle = 0, width=0.42\textwidth]{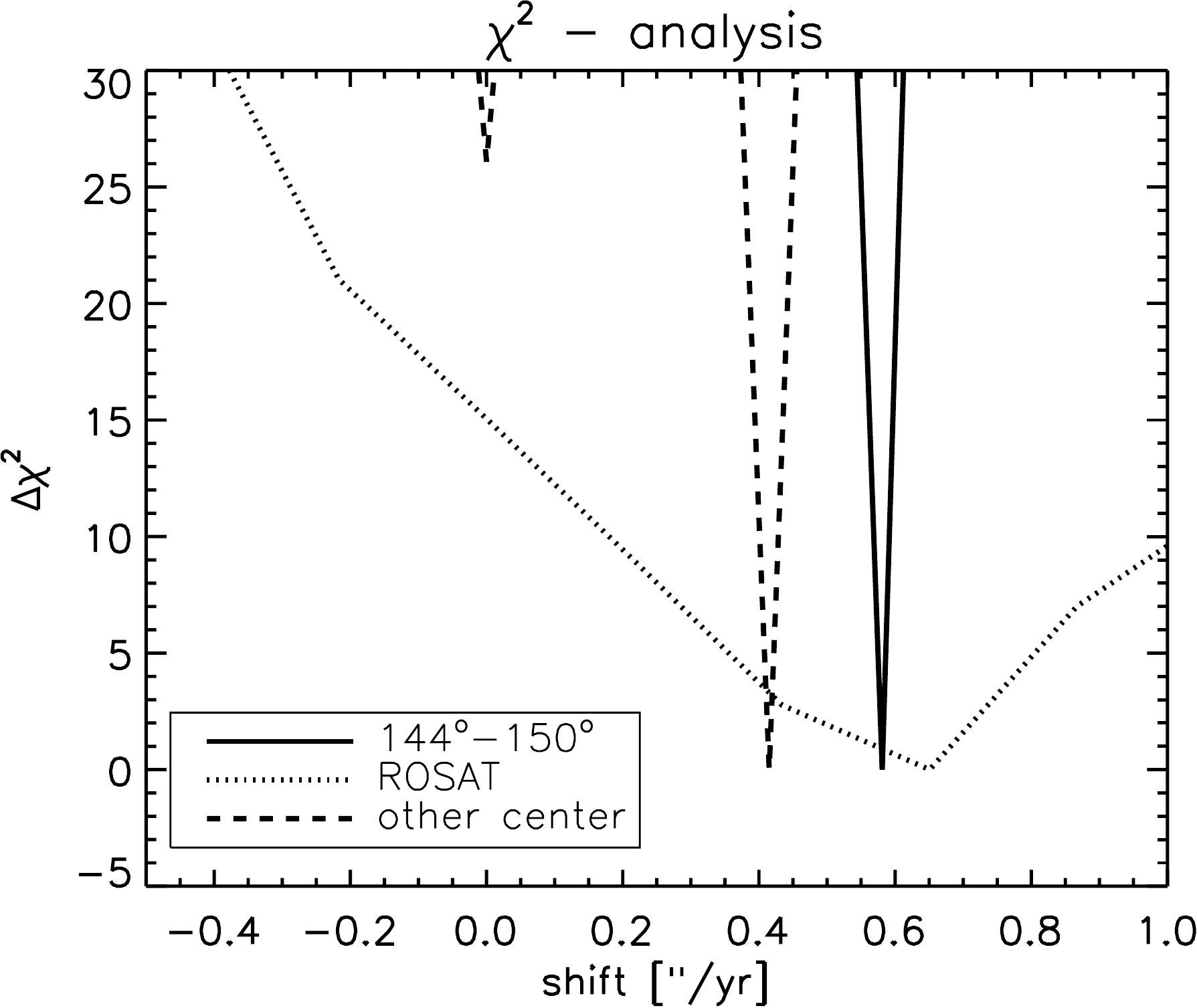} \\
\end{tabular}
\caption{
({\bf Left}) Steep gradient in the radial profiles for the 2004 and 2007 observations, adaptively binned with the Haar method\cite{Starck}, so that each bin has a signal to noise ratio of at least 4, with a maximum bin width of 2$''$. ({\bf Right}) $\chi^2$ statistics of the proper motion measurement. (Details of the used radial profiles in SOM).
}
             \label{ghi}%
\end{center}
\end{figure*}

We measured the proper motion of the shock at the location of the slit of the \halpha\ spectrum by comparing the positions of the shock in the two images (see SOM for further details). A solid estimate of the proper motion is $1.5\pm0.5 \arcsec$ in 3 years time (Fig. 3 and S1), implying a shock velocity  of $(6.0 \pm 2.0) \times 10^3$ km/s at a distance of 2.5 kpc \cite{Rosado,Sollerman}. The statistical error on the measured expansion is of the order of 0.2$\arcsec$. However, in the process of calculating the proper motions, we found that small details, such as slightly changing the angle in which we made the profile, tended to give a different proper motion, with a difference larger than the 0.2$\arcsec$~ statistical error we measured. However, in none of the measurements, did we find a  proper motion below 1.0$\arcsec$. Because the proper motion is higher than expected \cite{Vink2006}, we verified that it is consistent with data taken in 1993 with the Position Sensitive Proportional Counter (PSPC) on board the ROentgen SATellite (ROSAT) compared with the 2007 observation (Fig. 3 and S1). Although the proper motion, using the nominal pointing of the ROSAT PSPC, is statistically highly significant, the large pointing error of ROSAT ($\sim$4$''$) results in a detection of the proper motion at the $2\sigma$ level.


Compared to other remnants of a similar age, the shock velocity is surprisingly high. Recent models \cite{Dwarkadas2005} predict $v_{\rm s}\sim 5000$ \kms\ after 2000 years for SNRs evolving in a wind blown bubble \cite{effect}. This fits with the scenario where RCW~86 is evolving in a cavity and the southwest corner, which has a slower shock velocity \cite{Long,Ghavamian2001} and a mostly thermal \cite{Rho2002} X-ray spectrum, has already hit the cavity shell. 
Shock acceleration theory suggests that only shocks with velocities exceeding 2000 \kms\ emit X-ray synchrotron emission\cite{Aharonian1999,Vink2006}, which is also consistent with observations \cite{Katsuda2008}.
 
An additional uncertainty in the shock velocity is in the distance to RCW~86, which is based on converging but indirect lines of evidence. RCW~86 was found to be in the same direction as an OB association, at a distance of 2.5 kpc\cite{Westerlund}. Because high mass stars are often found in such associations, the progenitor of RCW~86 may well have formed in this one, provided that RCW~86 is the remnant of an exploded massive star. Other studies \cite{Rosado,Sollerman} found a distance of 2.3 and 2.8 kpc respectively, based on the line-of-sight velocity of ISM swept up by the remnant, combined with an observationally determined rotation curve of the Galaxy \cite{Brand}. The third argument supporting a distance of 2.5 kpc is the molecular supershell seen in CO emission in the direction of RCW~86, whose line-of-sight velocity agrees with that of RCW~86\cite{Matsunaga}. In further calculations, we take the distance towards RCW~86 to be 2.5 $\pm$~0.5 kpc, leading to a shock velocity of $6000 \pm 2800$~km/s. 

The relation between shock velocity and measured post-shock proton temperature has been extensively studied \cite{Chevalier1980,Ghavamian2001,Heng2007,Adelsberg,Ghavamian2007}, including the cross sections for excitation and charge exchange as function of $v_{\rm s}$. Although recent studies show that there can be a substantial effect of cosmic rays on the post-shock proton spectrum \cite{Raymond2008}, up to now, there was no need to include cosmic-ray acceleration in the interpretation of the post-shock temperature. 
This is possibly because most of the \halpha\ spectra are taken from the
brightest rims of SNRs. Because \halpha\ emission and efficient cosmic-ray acceleration are likely to anti-correlate \cite{Drury1996}, these rims probably have low cosmic-ray acceleration efficiency. A possible exception is `knot g' in the Tycho SNR, where indications for cosmic-ray acceleration in the form of a precursor have been found \cite{Ghavamian2000,Lee2007,Wagner}. 
Additionally, for some SNRs \cite{Winkler2003,Smith1991} the distance has been determined using the post-shock proton temperature in combination with the proper motion, using theoretical models which do not take into account energy losses and cosmic-ray pressure. This procedure leads to an underestimate of the distance if cosmic-ray acceleration is present. Thus, unless the distance is accurately determined in an independent way, there will be no discrepancy between the predicted $v_{\rm s}$, based on $kT$ and Eq. \ref{heating} and the actual shock velocity.

\begin{figure*}[!t]
\begin{center}
\includegraphics[angle = 0, width=0.7\textwidth]{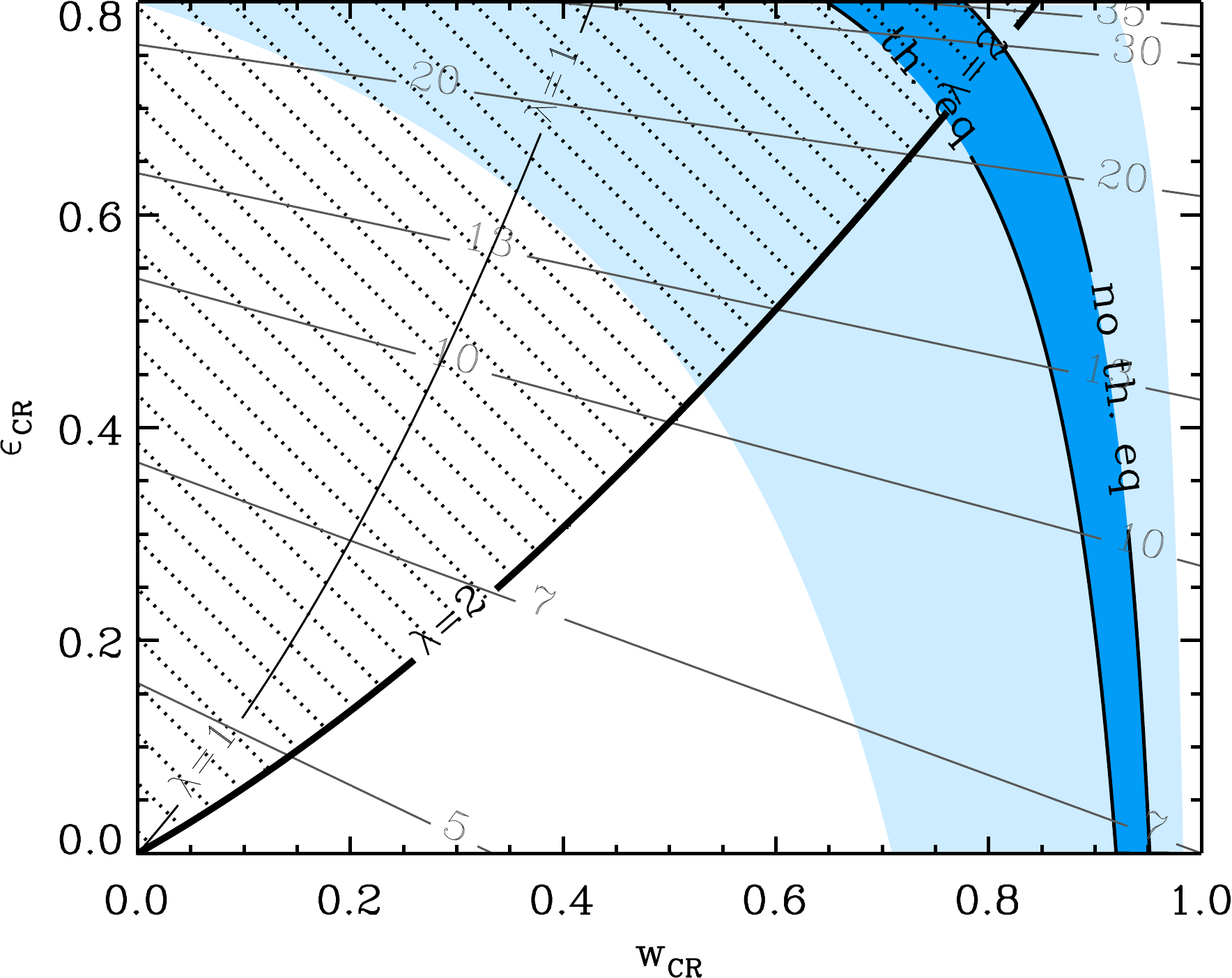} \\
\caption{
The dark blue area is the area allowed according to the modified equations, limited by full thermal equilibrium and no thermal equilibrium for the nominal values of $kT = $2.3 keV, $v_{\rm s}=$ 6000 km/s. The light blue area shows the area of the $kT$ and $v_{\rm s}$ with all uncertainties taken into account. 
The thin lines indicate the compression ratio ($\chi$) of the post-shock plasma. 
Within the allowed region, $\lambda = 2$ line provides an upper limit on $\epsilon_{\rm CR}$ and a lower limit on $w_{\rm CR}$.
			}
              \label{energy}%
\end{center}
\end{figure*}

The shock velocity of the X-ray synchrotron rim implies a post-shock temperature of 70 keV (assuming no thermal equilibrium), 42 keV (assuming equilibrium), whereas the measured post-shock temperature is 2.3 keV. This measurement is at least a factor 18 less than the post-shock temperature estimated from the shock velocity, which can now be used to constrain current theoretical shock heating models \cite{Drury2009,Patnaude2009b}.  Additionally, this proton temperature is close to the electron temperature at the same location\cite{Vink2006}, implying fast thermal equilibration between both species, breaking the trend between the shock velocity and the measure of thermal equilibrium seen in previous observations\cite{Ghavamian2007,Adelsberg}.

To translate this discrepancy into the energy and pressure in cosmic rays, we followed the approach of \cite{Vinkreview2008}, which is based on standard shock equations for plane-parallel, steady-state shocks, modified by additional pressure and loss terms [see also \cite{Chevalier1983,Berezhko1999,Bykov2008}]. The loss term is defined in terms of the incoming energy flux: $\epsilon_{\rm CR} \equiv F_{\rm CR}/\frac{1}{2}\rho_0v_{\rm s}^3$, $F_{\rm CR}$ is the amount of energy flux in cosmic rays which escapes from the system and $\rho_0$ is the pre-shock density. The parameter which indicates the fraction of the pressure induced by cosmic rays in the total post-shock pressure is $w_{\rm CR}$ ($w_{\rm CR}\equiv P_{\rm NT}/(P_{\rm T} + P_{\rm NT})$, with $P_{\rm T}$ the pressure in particles with a thermal and $P_{\rm NT}$ with a non-thermal energy distribution (i.e. CRs). We plot the modified equations (listed in the supporting online material) in Figure 4 and indicate the region where the combination of $kT$ and $v_s$ of the northern rim of RCW~86 resides for thermal equilibrium as well as for no thermal equilibrium. As Figure 4 shows, a post-shock temperature and a shock velocity do not give a unique solution for $w_ {\rm CR}$ and $\epsilon_{\rm CR}$. However, the cosmic rays significantly change the shock dynamics, because the combination solution is far away from $w_{\rm CR}= 0$ and $\epsilon_{\rm CR}= 0$ (Fig. 4).

There are two ways to further constrain $w_{\rm CR}$ and $\epsilon_{\rm CR}$. First, an additional estimate of the compression ratio ($\chi$) of the post-shock plasma would exactly determine $w_ {\rm CR}$ and $\epsilon_{\rm CR}$. For certain SNRs this is done by determining the distance between the supernova ejecta and the outer shock; a higher compression ratio implies that the swept-up ISM forms a thinner shell and hence the ejecta will be closer to the shock front \cite{Warren2005,Cassam2008}. However, ejecta and swept-up ISM are only distinguished by their thermal spectra, which is (almost) absent in the X-ray synchrotron dominated rim ($\sim 15\%$ of the total X-ray emission \cite{Vink2006}). 

An other way is to invoke a dependency of  $w_ {\rm CR}$ on $\epsilon_{\rm CR}$. According to non-linear shock acceleration theory\cite{Malkov,Drury2009}, $\epsilon_{\rm CR}/w_{\rm CR} = \frac{2}{\lambda}(1-1/\chi)^2$, in which $\chi$ is the compression ratio of the post-shock plasma and  $\lambda = 1,2$ indicates the $(w_{\rm CR},\epsilon_{\rm CR})$ relation for a cosmic-ray spectrum with $f(p)\propto p^{-3.},p^{-3.5}$ respectively, with $p$ the momentum of the cosmic rays. The $\lambda=2$ line gives an upper limit to the energy losses, since it is valid for the most efficient cosmic-ray acceleration by cosmic-ray modified shocks \cite{Malkov1999}. For $f(p)\propto p^{-4}$, $\lambda = ln(p_{\rm max}/mc)$\cite{Drury2009} which can be large and does not provide a lower limit to $\epsilon_{\rm CR}$. Taking the $\lambda = 2$ line as an upper limit for $\epsilon_{\rm CR}$, we find a value for $w_{\rm CR}$ of $\ge 50$\%. One remaining question is whether we should include the effects of the turbulent magnetic field. The average magnetic field pressure in RCW~86 has been estimated to be $P_B=B^2/8\pi=2.3\times10^{-11}$dyn cm$^{-2}$, for a magnetic field of 24 $\mu$G \cite{Vink2006}. This is an order of magnitude below the thermal pressure, which we estimate to be $P_T = nkT=3.7\times10^{-10}$dyn cm$^{-2}$, for n = 0.1\cite{Vink2006} and kT = 2.3 keV.  In reality, the magnetic field pressure may be higher if one takes full account of its unknown, turbulent spectrum.

In summary, our observations show that the post-shock temperature of the northeast rim of RCW~86 is lower than expected from standard shock relations using the measured shock velocity. 
The high velocity ($6000 \pm 2800$ km/s) of the shock implies a local low ISM density, which can be expected in a cavity blown by a stellar wind. Cosmic-ray acceleration decreases the post-shock proton temperature in RCW~86 by a factor of 18, implying that $\ge 50$\% of the post-shock pressure is due to cosmic rays.

\bibliography{1173383Revisedtext.bib}

\bibliographystyle{Science}


\begin{scilastnote}
\item 
We wish to thank A. Achterberg for a useful discussion on the current literature on the theory of shock heating. EH and JV are supported by the Vidi grant of JV from the Netherlands Organization for Scientific Research (NWO). This work was supported in part by Grant-in-Aid for Scientific Research of the Japanese Ministry of Education, Culture, Sports, Science and Technology, no. 19á4014 (AB) and no. 19047004 and no. 21740184 (RY). SF is supported by SAO grant G07-8073X. PG is supported by the STScI grant GO-11184.07. This paper is based in part on observations made with ESO Telescopes at the Paranal Observatories under programme ID 079.D-0735. 
\end{scilastnote}


\clearpage


\clearpage
\begin{appendix}
\section{Supporting online material}
\paragraph*{VLT spectra}
\begin{table*}[t!]
\begin{center}
\caption{ Journal of the VLT long slit spectroscopic observations}\label{VLT_journal}
\smallskip
\begin{tabular}{c|c|c|c|c|c}\noalign{\smallskip}
Pointing &  $\alpha^1$& $\delta^1$ & observation date & exposure time & position angle \\ 
& (J2000) & (J2000) & & & (East of North) 
{\smallskip}\\\hline\noalign{\smallskip}
NE & 14:45:15.7 & -62:16:33.2 & 05/16/2007 & 2734 s & 240$^{\circ}$\\ 
& 14:45:15.7 & -62:16:33.8 & 07/16/2007 & 2734 s & 240$^{\circ}$\\
 & 14:45:15.7 & -62:16:34.6 & 07/18/2007 & 2734 s & 240$^{\circ}$\\ 
 & 14:45:15.7 & -62:16:34.0 & 07/20/2007 & 2734 s & 240$^{\circ}$ \\
 \noalign{\smallskip}
\end{tabular}
\\$^{1}$Coordinates given are the center of the CCD with which we took the spectrum. The slit, which was not in the middle of the CCD, was laid over the filament with coordinates $\alpha =$ 14:45:02.813, $\delta = $-62:16:33.05 (J 2000).
\end{center}
\end{table*}

Table \ref{VLT_journal} lists observational characteristics of the VLT observation. To reduce the data, we first calculated for each pixel the median signal in a square of 5$\times$5 pixels around it. Then, the pixels which had a higher value than 2$\times$ the median were flagged as cosmic-ray pixels, and are not used in the following calculations. We removed the skylines by fitting a 3$^{\rm rd}$ order polynomial\footnote{Or a lower order, depending on whether the fit improved by taking the higher order. } to the spectrum in the spatial direction for each wavelength coordinate. Furthermore, we calibrated the wavelength using the HeHgCdArNe calibration spectrum. Then, we added the spectra of the 4 observations, resulting in the spectrum shown in Figure \ref{Spectra}. We fitted this spectrum by minimizing the $\chi^2$, the line profile using two Gaussians, convolved with a hat profile with the width of the slit  (2.5$''$). The errors on the data points are based on the variance of the residuals between the data and the best fit model. This method results by definition in a $\chi^2/{\rm d.o.f.} \sim 1$. The brightness of the filaments is $1.0 \pm 0.2 \times 10^{-16} \rm{erg~s^{-1}~cm^{-2}~arcsec^{-2}}$.

\paragraph*{Expansion measurement \& statistics}

\begin{table*}[h!]
\begin{center}
\caption{\label{Chandra_journal} Journal of the Chandra observations}
\smallskip
\begin{tabular}{c|c|c|c|c|c}\noalign{\smallskip}
ObsID &  $\alpha$ & $\delta$ & observation start date & exposure time & Roll angle\\
 & (J2000) & (J2000) & &&
{\smallskip}\\\hline\noalign{\smallskip}
4611 & 14:45:03.60 & -62:21:05.56 & 06/15/2004 &  69.1 ks & 295.16 \\ 
7642 & 14:45:04.48 & -62:20:40.53 & 06/20/2007 &  71.7 ks & 299.02 \\ 
\noalign{\smallskip}

\end{tabular}
\end{center}
\end{table*}

Table \ref{Chandra_journal} lists the observational parameters of both Chandra observations.  The data were reduced using the CIAO data reduction package, version 4.0 and the calibration database CALDB 3.4.
We turned off the randomization, which is applied during the standard process of generating the event list; combined with the dithering of the telescope this gives a slightly better angular resolution.
We checked the registering of the two observations using 9 point sources in the field. The error in the alignment of both pointings is well below 0.1$\arcsec$, which is the value we are using as a systematic error in the proper-motion measurement. From the event list, we made radial surface brightness profiles with bins in the radial direction of 0.25$''$, using photons with energies between 0.5 and 6.0 keV. We choose the center of the proper motion to be $\alpha= 14^{\rm h}42^{\rm m} 31.00^{\rm s}$ and $\delta = -62^{\circ}29'34.99''$ (J 2000). This is not necessarily the center of the remnant, it is our estimated center of the curvature of the part of the remnant we are interested in. Note that a wrongly chosen center can only result in a lower proper motion of the shock. 
We calculated the expansion in 2 overlapping regions (Figure \ref{Image}). We implemented the Poisson statistics as follows: for each bin, we calculated the probability that the number of counts in both bins were drawn from the same Poisson distribution. 
We measure the proper motion by shifting the normalized profiles with respect to each other, with steps of 1 bin and for each shift, we calculate the probability that both profiles were drawn from the same distribution, using the $\chi^2$ and the Poisson maximum likelihood method. Additionally, we used the Kolmogorov Smirnov statistic. For applying the latter, we first made cumulative distributions for the profiles and then calculated the Kolmogorov Smirnov statistic for each shift.

The Kolmogorov-Smirnov statistic does not provide an error on the parameters. To estimate the errors, we simulated 20 radial profiles using the bootstrap method \cite{numerical}, added an artificial proper motion (1.0$\arcsec$) and measured this proper motion using the Kolmogorov-Smirnov statistic. The best fit  proper motion agrees well with the input value. We use the standard deviation on these best fit proper motions as the 1$\sigma$ error. In addition, we checked the Poison maximum likelihood estimation and $\chi^2$ statistics in a similar way, resulting in consistent estimates of the artificial proper motion (1.0$\arcsec$ $\pm$ 0.2$\arcsec$). Furthermore, we use these simuations to determine the optimum choice for the range in radius. The most reliable results for the proper motion were obtained if we include the whole rim in the radial profile. We validated our result using two independently written computer codes.

\begin{figure*}[!h]
\begin{center}
\begin{tabular}{c c c}
\includegraphics[angle = 0, width=0.4\textwidth]{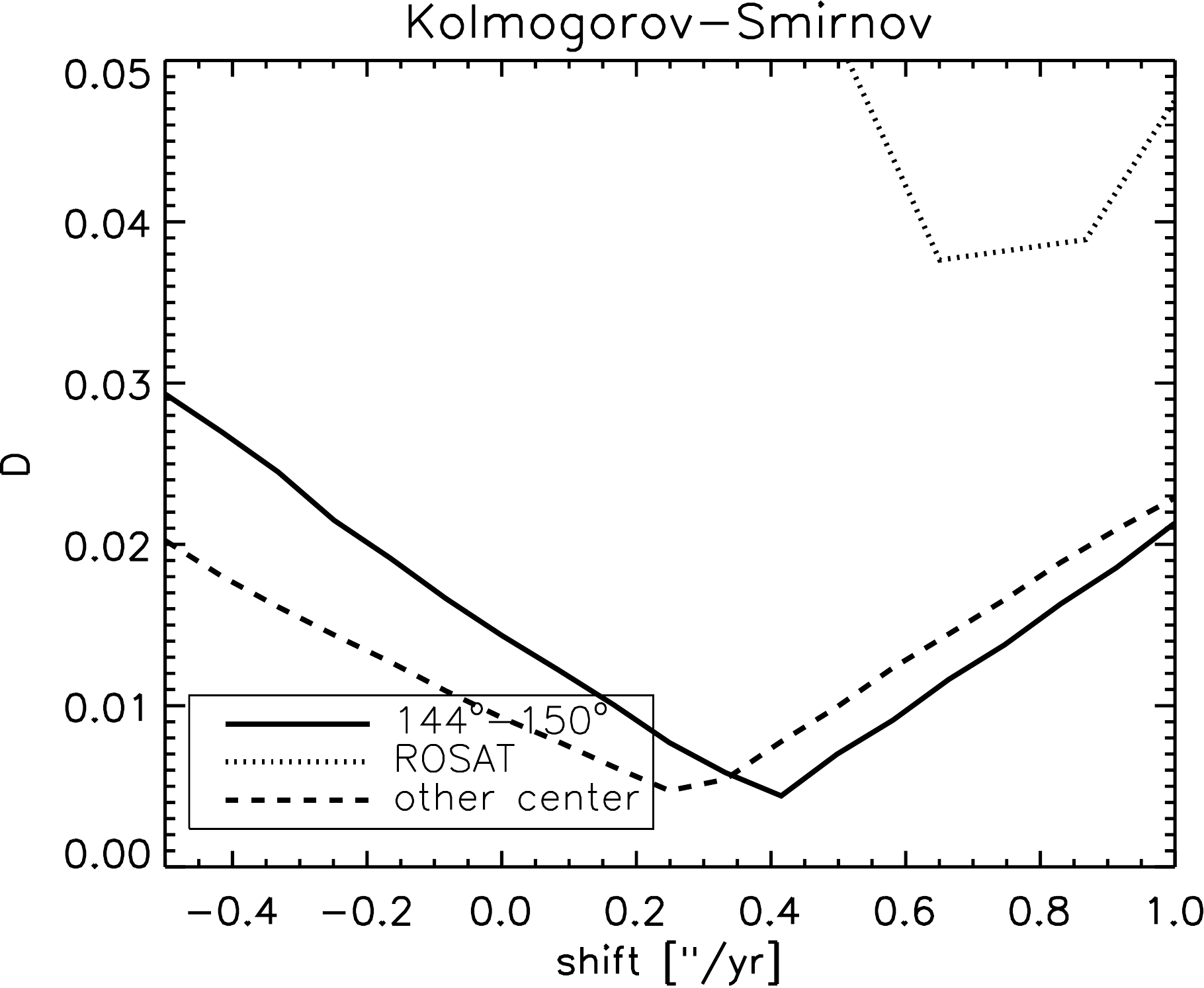} &
\includegraphics[angle = 0, width=0.39\textwidth]{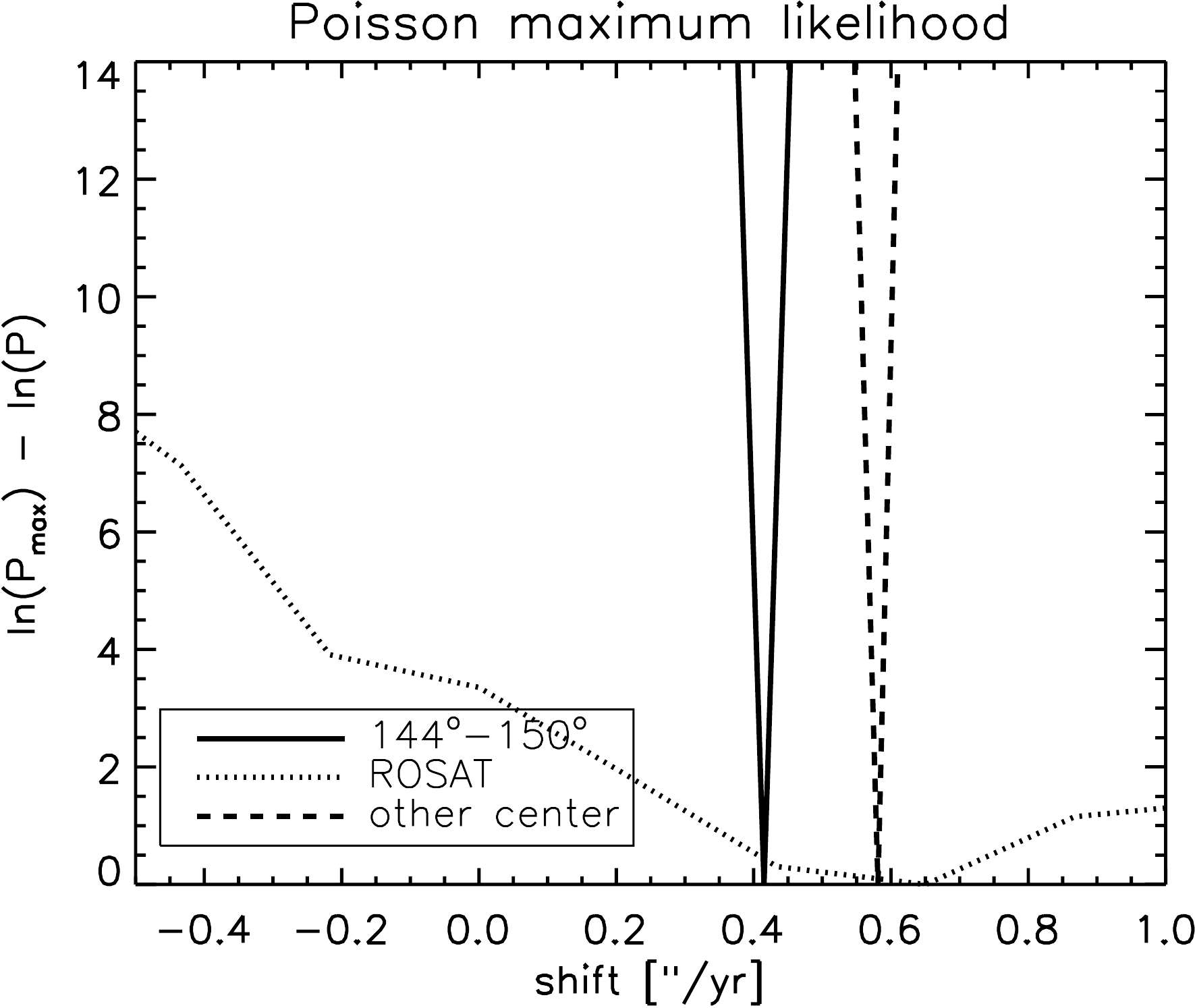}  \\
\end{tabular}
\caption{Plotted are the statistics on the proper motion of RCW~86, for both the Kolmogorov Smirnov statistic and the Poisson maximum likelihood estimation. The linestyles denote the same regions as in Figure \ref{ghi}. Since the Poisson statistics is a multiplication of the probabilities for each bin, the total probability depends on the number of bins in the profile. We corrected for that by subtracting $ln(P_{\rm max})$, we multiplied the y-axis with -1, so the best fit solution is a minimum in this plot. }
              \label{proper}%
\end{center}
\end{figure*}

\paragraph*{kT and V$_{\rm s}$ relation}
Analytical equations of the post-shock temperature and shock velocity in the presence of cosmic-ray acceleration have been described by several authors \cite{Chevalier1983, Berezhko1999,Blasi}. In this section, we give a summary of the equations used to obtain Figure \ref{energy}. We start out with a relation which states the conservation of momentum over the shock front:
\begin{equation}
P_2+\rho_2 u_2^2 = P_0 + \rho_0u_0^2.
\end{equation}
A '0' subscript means pre-shock and a '2' subscript means post-shock, $\rho$ denotes the density, $P$ the pressure and $u$ the velocity of the gas/plasma in the frame of the shock. Now, we use conservation of mass over the shock: $\rho_0 u_0 = \rho_2 u_2$ and we define the compression ratio $\chi \equiv \rho_2/\rho_0$:
\begin{equation}
P_2 = P_0 + \rho_0 u_0^2(1-1/\chi).
\end{equation}
We introduce $w_{\rm CR}\equiv P_{\rm NT}/(P_{\rm T} + P_{\rm NT})\rightarrow P_2 = P_{\rm T} /(1-w_{\rm CR})$. Additionally, we use $P_T = nkT_i$, $\rho = n m_i$ and we assume that the pre-shock pressure is small compared to $\rho_0 u_0^2(1-1/\chi)$:
\begin{equation}
 kT_i =  (1-w_{\rm CR})\frac{1}{\chi}(1-1/\chi)m_i u_0^2.
\end{equation}
Note that for a shock without cosmic-ray pressure ($w_{\rm CR}=0$) and no energy losses and a non-relativisitic gas ($\chi = 4$), we get equation \ref{heating}. To derive the compression ratio $\chi$, we use $\gamma_s$; the effective adiabatic index at the shock front. This is defined as \cite{Chevalier1983}:
\begin{equation} 
\gamma_s = \frac{5+3w_{\rm CR}}{3(1+w_{\rm CR})}\label{gamma_s},
\end{equation}
We now use $\gamma_s$ in the equation for compression ratio including energy losses by cosmic rays, as described in \cite{Bykov2008} and define $G = \frac{3}{2}w_{\rm CR} + \frac{5}{2}$. This gives:
\begin{equation}
\chi  = \frac{G+\sqrt{G^2-(1-\epsilon_{\rm CR})(2G-1)}}{1-\epsilon_{\rm CR}}.
\end{equation}

\end{appendix}


\clearpage


\end{document}